\documentstyle[preprint,aps,floats,epsf]{revtex}
\newcommand{\tr}{\,{\rm tr}}
\newcommand{\vp}{\varphi}

\newcommand{\hD}{\hat{D}}
\newcommand{\hvp}{\hat{\varphi}}
\newcommand{\pd}{{\partial} }

\begin{document}
\setcounter{footnote}{1}
\draft
\title{The Chern-Simons Coefficient in Supersymmetric Non-Abelian Chern-Simons Higgs Theories}
\author{Hsien-chung Kao\footnote{Email address:hckao@mail.tku.edu.tw}}
\address{Department of Physics, Tamkang University, Tamsui, Taiwan 25137, R.O.C.}


\maketitle
\begin{abstract}
By taking into account the effect of the would be Chern-Simons
term, we  calculate the quantum correction to the Chern-Simons
coefficient in supersymmetric Chern-Simons Higgs theories with
matter fields in the fundamental representation of $SU(n)$.
Because of supersymmetry, the corrections in the symmetric and
Higgs phases are identical. In particular, the correction is
vanishing for $N=3$ supersymmetric Chern-Simons Higgs theories.
The result should be quite general, and have important implication
for the more interesting case when the Higgs is in the adjoint
representation.
\end{abstract}

Chern-Simons theories can give rise to particle excitations with
fractional spin and statistics, and thus have been used as
effective field theories to study the fractional quantum Hall
effect \cite{CSFT,CSANYON,FQHE}.  They are also interesting when
the Higgs fields with a special sixth order potential are included
so that the systems admit a Bogomol'nyi bound in energy
\cite{Bogo}. The bound is saturated by solutions satisfying a set
of first-order self-duality equations \cite{Hong}. These solutions
have rich structure and have been under extensive study especially
when the gauge symmetry is non-abelian with the Higgs in the
adjoint representation \cite{NASD}.  It is known that the
self-duality in these systems signifies an underlying $N=2$
supersymmetry and thus the Bogomol'nyi bound is expected to be
preserved in the quantum regime \cite{CSII}. Furthermore, when
these theories are dimensionally reduced, an additional Noether
charge appears, which in turns yields a BPS-type of domain wall
\cite{Domn}.

The quantum correction to the Chern-Simons coefficient has also
attracted a lot of attention.  For theories without  massless
charged particles and the gauge symmetry is not spontaneously
broken, Coleman and Hill have shown in the abelian case that only
the fermion one-loop diagram can contribute to the correction to
the Chern-Simons coefficient and yields ${1\over 4\pi}$
\cite{Coleman}.  The quantization of the correction can be
understood with an topological argument in the spinor space by
making use of the Ward-Takahashi identity \cite{Ishikawa}. When
there is spontaneous breaking of gauge symmetry, one can show that
there exists in the effective action the so-called would be
Chern-Simons terms, which induces terms similar to the
Chern-Simons one in the Higgs phase \cite{Khleb2}.  By taking into
account the effect of the would be Chern-Simons term, it has been
shown that the one-loop correction in the Higgs phase is identical
to that in the symmetric phase \cite{CHTh}.  On the other hand, if
the charged particles, both scalars and spinors can
contribute to the correction at two-loop level and it is not
quantized \cite{Semenoff}.

The situation becomes even more intriguing when the gauge symmetry
is non-abelian: the Chern-Simons coefficient must be integer
multiple of ${1\over 4\pi}$ for the systems to be invariant under
large gauge transformation; otherwise the theories are not
quantum-mechanically consistent.  Therefore, it is interesting to
confirm that the quantization condition is not spoiled by quantum
effects.  In the symmetric phase, this has been shown to one loop
\cite{Pisarski}.  When there is no bare Chern-Simons term, it is
also verified up to two loops considering only the fermionic
contribution \cite{YCKao}.  In the Higgs phase, it is known for
some time that if there is remaining symmetry e.g. $SU(n)$ with
$n\ge 3$,  the quantization condition will still be satisfied
\cite{Dunne1,Khare,PV}.  However, if the gauge symmetry is
completely broken, e.g. $SU(2)$, simple-minded calculation shows
that the correction is again complicated and not quantized
\cite{Khleb2}. Although one may argue that this arises because
there is no well-defined symmetry generator in such case, a better
way to understand the whole thing is again to note the effect of
the would be Chern-Simons terms. They are invariant even under the
large gauge transformation, and their coefficients need not to be quantized. Therefore, we must subtract out the their contribution to obtain the correct result. Indeed, more careful calculation shows that for the Higgs being in fundamental
$SU(n)$ the quantization condition is always satisfied whether the
gauge symmetry is completely broken or not \cite{Nahill}.  As a
result, a more or less unifying picture of the quantum correction
to the Chern-Simons coefficient has emerged.

In pure non-abelian Chern-Simons theories, there is also the so-called regularization dependence of the quantum corrections to the Chern-Simons coefficient: 
$$
\Delta \kappa = {\rm sign}(C_v), 
$$
if we introduce the Yang-Mills term as a UV regulator, while
$$
\Delta \kappa = 0,
$$
if we do not \cite{RegDep}. Here, $C_v$ is the the quadratic Casimir operator in the adjoint representation of the gauge group.  Further studies suggest that every local regulator manifestly preserving BRS invariance and unitarity would give rise to the same quantum correction \cite{RuizI}.  Interestingly, it has been shown that $N=1$ supersymmetric Yang-Mills-Chern-Simons theory is finite to all orders. \cite{RuizII} Moreover, if the regulator is supersymmetric, the corrections become regularization independent \cite{SUYMCS}.  In particular, the corrections are vanishing for $N=2,3$ supersymmetric Chern-Simons theories.  Hence, we would like to know what happen if there is also spontaneous breaking of gauge symmetry in the system.

In this paper, we calculate the quantum corrections to the
Chern-Simons coefficient in supersymmetric Chern-Simons Higgs
theories with the Higgs being in the fundamental $SU(n)$.  It
turns out that the result is partially regularization dependent.
If we do not introduce the Yang-Mills term, the quantum
corrections are quantized and identical in the symmetric and Higgs
phase because of supersymmetry.  On the other hand, if we do, the
result is more complicated.  For $n\ge 3$, the quantum corrections
are still identical in the two phases.  For $n = 2$, however, the
quantum corrections becomes different in the two phases.  We
conclude with some comments on its implication and possible future
direction.

With matter fields in the fundamental $SU(n),$ the $N=3$
supersymmetric nonabelian Chern-Simons Higgs theories can be
simplified to \cite{SUSY}:
\begin{eqnarray}
{\cal L} = &\;& - i\kappa \; \epsilon^{\mu\nu\rho} \tr \biggl\{
A_\mu \partial_\nu A_\rho - {2\over 3}iA_\mu A_\nu A_\rho \biggr\}  + |D_\mu \phi_1|^2 + |D_\mu \Phi_2|^2
+ \bar{\psi}\gamma^\mu D_\mu \psi + \bar{\chi}\gamma^\mu D_\mu \chi \nonumber \\
&\;& + {1\over \kappa^2}(|\phi_1|^2+|\Phi_2|^2)\left\{
\left[{(n-1)\over 2n}(|\phi_1|^2-|\Phi_2|^2)+ v^2 \right]^2
+ {1\over 4}|\phi_1|^2|\Phi_2|^2 + {(3n-2)(n-2)\over 4n^2}|\phi^\dagger_1 \Phi_2|^2 \right\} \nonumber \\
&\;& +{1 \over \kappa}\left\{
[v^2-{1\over 2n}|\phi_1|^2-{(2n-1)\over 2n}|\Phi_2|^2]\bar{\psi}\psi
+[-v^2-{(2n-1)\over 2n}|\phi_1|^2-{1\over 2n}|\Phi_2|^2]\bar{\chi}\chi \right\} \nonumber \\
&\;&+ {1\over 2}[ (\bar{\psi}\phi_1)(\phi^\dagger_1 \psi)
+ (\bar{\chi}\Phi_2)(\Phi^\dagger_2 \chi)]
-{(n-2)\over 2n}[ (\bar{\psi}\Phi_2)(\Phi^\dagger_2 \psi)
+ (\bar{\chi}\phi_1)(\phi^\dagger_1 \chi)]  \label{Lg} \\
&\;&+ {(n-1)\over 2n}[ (\bar{\psi}\phi_1)(\psi^\dagger \phi_1)
+ (\phi^\dagger_1 \bar{\psi}^\dagger)(\phi^\dagger_1 \psi)
+(\bar{\chi}\Phi_2)(\chi^\dagger \Phi_2)
+ (\Phi^\dagger_2\bar{\chi}^\dagger)(\Phi^\dagger_2 \chi)] \nonumber \\
&\;&-{(n-1)\over 2n}[ (\Phi^\dagger_2\phi_1)(\bar{\psi}\chi)
+ (\phi^\dagger_1 \Phi_2)(\bar{\chi}\psi)
+(\bar{\psi}\phi_1)(\Phi^\dagger_2 \chi)
+ (\bar{\chi}\Phi_2)(\Phi^\dagger_2 \chi)] \nonumber \\
&\;&-[ (\bar{\psi}\phi_1)(\chi^\dagger \Phi_2) + (\Phi^\dagger_2\bar{\chi}^\dagger)(\phi^\dagger_1 \psi)]
+{1\over n}[ (\bar{\psi}\Phi_2)(\chi^\dagger \phi_1) + (\phi^\dagger_1\bar{\chi}^\dagger)(\Phi^\dagger_2 \psi)]. \nonumber
\end{eqnarray}
Here $D_\mu = (\partial_\mu - i A^m_\mu T^m)$ and $\gamma_\mu=
\sigma_\mu$ so that the gamma matrices satisfy $\gamma_\mu
\gamma_\nu = \delta_{\mu\nu} + i\epsilon_{\mu\nu\rho}
\gamma_\rho$, with $\epsilon_{012} = 1$. The generators satisfy
$[T^m, T^n] = if^{lmn}T^l,$ with the normalization $\tr\{T^m T^n\}
= \delta^{mn}/2$ and $\sum_m
(T^m)_{\alpha\beta}(T^m)_{\gamma\delta} ={1\over
2}\delta_{\alpha\delta} \delta_{\beta\gamma} - {1\over
2n}\delta_{\alpha\beta} \delta_{\gamma\delta}.$

We will use the background field gauge so that the effective
action is explicitly gauge invariant and the gauge fields do not
get renormalized.  This can be done by separating $A_\mu$ into the
background part $A_\mu$ and the quantum part $Q_\mu.$  In the
Higgs phase, $\Phi_2 = \phi_2 + \vp$ with $\vp^\dagger\vp =
|\vp|^2$.  As usual, the gauge fixing and the Faddeev-Popov ghost
terms are given by
\begin{equation}
{\cal L}_{gf}={1\over 2\xi}\biggl\{\bigl(\hD_\mu Q_\mu\bigr)^m
+ i\xi\bigl(\vp^\dagger T^m\phi_2 - \phi_2^\dagger T^m\vp \bigr) \biggr\}^2,
\label{Lgf}
\end{equation}
and \begin{eqnarray}
{\cal L}_{FP}  =
&\;& 2\,\tr \biggl\{\bigl(\hD_\mu \bar{\eta} \bigr) \bigl(\hD_\mu \eta\bigr)
-i\bigl(\hD_{\mu} \bar{\eta} \bigr) \bigl[Q_{\mu}, \eta \bigr] \biggr\}\nonumber\\
&\;& + \xi\bigl(\vp^\dagger\bar{\eta}\eta\vp - \vp^\dagger\eta\bar{\eta}\vp\bigr)
+ \xi\bigl(\vp^\dagger\bar{\eta}\eta\phi_2 - \phi_2^\dagger\eta\bar{\eta}\vp\bigr).
\label{Lgh}
\end{eqnarray}
Here $\hD_\mu$ is the covariant derivative using the background
field. Combining Eqs. (\ref{Lg}), (\ref{Lgf}) and (\ref{Lgh}), we
see the relevant quadratic terms are
\begin{eqnarray}
{\cal L}_0 =
&\;& {1\over 2} Q^m_\mu \biggl\{
\bigl[i\kappa \epsilon_{\mu\nu\rho} \pd_\rho \bigr] \delta_{mn}
- {1\over \xi}\pd_\mu \pd_\nu + \delta_{\mu\nu}\bigl[ (\vp^\dagger T^mT^n\vp)
+ (\vp^\dagger T^nT^m\vp)\bigr]\biggr\}Q^n_\nu \nonumber\\
+ &\;& {1\over 2} (\phi_2^\dagger, \phi_2^T) \left(
\matrix{ -\pd^2 + {(n-1)^2|\vp|^2\vp\vp^\dagger\over 2n^2\kappa^2}
+ {\xi|\vp|^2\over 2} - {\xi\vp\vp^\dagger\over 2n}
& {(n-1)^2|\vp|^2\vp\vp^T\over 2n^2\kappa^2}-{(n-1)\xi\vp\vp^T\over 2n} \cr
{(n-1)^2|\vp|^2\vp^*\vp^\dagger\over 2n^2\kappa^2} - {(n-1)\xi\vp\vp^\dagger\over 2n}
& -\pd^2 + {(n-1)^2|\vp|^2\vp^*\vp^T\over 2n^2\kappa^2} + {\xi|\vp|^2\over 2}
- {\xi\vp^*\vp^T\over 2n} }\right) \left( \matrix{\phi_2 \cr \phi_2^* }\right) \nonumber\\
+ &\;& {1\over 2} (\bar{\psi}, \bar{\psi}^*)
\left( \matrix{ \gamma\cdot \pd - {|\vp|^2\over 2\kappa} -
{(n-2)\vp\vp^\dagger\over 2n\kappa} & 0 \cr
0 & \gamma\cdot \pd - {|\vp|^2\over 2\kappa} - {(n-2)\vp^*\vp^T\over 2n\kappa} }\right)
\left( \matrix{\psi \cr \psi^*}\right) \nonumber \\
+ &\;& {1\over 2} (\bar{\chi}, \bar{\chi}^*)
\left( \matrix{ \gamma\cdot \pd - {|\vp|^2\over 2\kappa} + {\vp\vp^\dagger\over 2\kappa} & {(n-1)\vp\vp^T\over n\kappa} \cr
{(n-1)\vp^*\vp^\dagger\over n\kappa} & \gamma\cdot \pd - {|\vp|^2\over 2\kappa} +
{\vp^*\vp^T\over 2\kappa} }\right) \left( \matrix{ \chi \cr \chi^*
}\right) \label{LBq} \\
+ &\;& f^{lmn} \biggl\{{1\over \xi}(\pd_\mu Q^l_\mu) A^m_\mu Q^n_\nu - {i\kappa\over 2}
\epsilon_{\mu\nu\rho} A^l_\mu Q^m_\nu Q^n_\rho \biggr\} \nonumber\\
+ &\;& 2 (\vp^\dagger A_\mu Q_\mu \phi_2)
+ 2(\phi_2^\dagger Q_\mu A_\mu \vp)
-i \bar{\psi}\gamma^\mu A_\mu \psi -i \bar{\chi}\gamma^\mu A_\mu \chi. \nonumber
\end{eqnarray}

In our case, there are two relevant would be Chern-Simons terms:
\begin{eqnarray}
&\;& O_1 = \epsilon^{\mu\nu\rho}
i\bigl\{\Phi_2^\dagger T^m (D_\mu \Phi_2) -(D_\mu\Phi_2)^\dagger T^m \Phi_2 \bigr\}
F^m_{\nu\rho}, \nonumber\\
&\;& O_2 = \epsilon^{\mu\nu\rho}
i\bigl\{\Phi_2^\dagger (D_\mu \Phi_2) - (D_\mu\Phi_2)^\dagger \Phi_2 \bigr\}
(\Phi_2^\dagger F_{\nu\rho} \Phi_2).
\end{eqnarray}
In the Higgs phase, they give rise to
\begin{eqnarray}
&\;& \epsilon^{\mu\nu\rho} A^n_\mu F^m_{\nu\rho}
\bigl\{(\vp^\dagger T^m T^n\vp) + (\vp^\dagger T^n T^m\vp)\bigr\},
\nonumber\\
&\;& 2\epsilon^{\mu\nu\rho} A^n_\mu F^m_{\nu\rho}
(\vp^\dagger T^m\vp)(\vp^\dagger T^n \vp),
\label{WBCS}
\end{eqnarray}
respectively.  We note that their transformation property under
the $SU(n)$ symmetry are different from the quadratic part of the
Chern-Simons term.  Therefore, we will leave the VEV $\vp$ in
general form so that it is easier to extract the correction to the
Chern-Simons coefficient. For this purpose, we express the
propagators in terms of the following projection operators:
\begin{eqnarray}
&\;& (P_1)_{mn} = \delta_{mn} - 2 \bigl[(\hvp^\dagger T^mT^n \hvp)
+ (\hvp^\dagger T^nT^m \hvp) \bigr] + {2(n-2)\over (n-1)}
(\hvp^\dagger T^m \hvp) (\hvp^\dagger T^n \hvp), \nonumber\\
&\;& (P_2)_{mn} = 2 \bigl[(\hvp^\dagger T^mT^n \hvp) + (\hvp^\dagger
T^nT^m \hvp) \bigr] - 4 (\hvp^\dagger T^m \hvp) (\hvp^\dagger T^n \hvp), \nonumber\\
&\;& (P_3)_{mn} = {2n \over (n-1)}
(\hvp^\dagger T^m \hvp) (\hvp^\dagger T^n \hvp); \nonumber\\
&\;& Q_1 = \left( \matrix{ I - \hvp\hvp^\dagger & 0 \cr 0 & I - \hvp^*\hvp^T }\right), \nonumber\\
&\;& Q_2 = {1\over 2}\left(
\matrix{ \hvp\hvp^\dagger & \hvp\hvp^T \cr \hvp^*\hvp^\dagger & \hvp^*\hvp^T}\right),
\nonumber\\
&\;& Q_3 = {1\over 2}\left( \matrix{ \hvp\hvp^\dagger & -\hvp\hvp^T \cr
 -\hvp^*\hvp^\dagger & \hvp^*\hvp^T }\right),
\end{eqnarray}
where $\hvp \equiv \vp/|\vp|.$

With these projection operators, it is now straightforward to
obtain the propagators of $Q_\mu, \phi_2, \psi,$ and $\chi$:
\begin{eqnarray}
\Delta^{mn}_{\mu\nu}(k) = \biggl\{
&\;& \bigl[\Delta^1_{\mu\nu}(k) \bigr] (P_1)_{mn}
+ \bigl[\Delta^2_{\mu\nu}(k) \bigr] (P_2)_{mn}
+ \bigl[\Delta^3_{\mu\nu}(k) \bigr] (P_3)_{mn} \biggr\}, \nonumber\\
D(k) = \biggl\{
&\;& \bigl[ D^1(k)\bigr] Q_1
+ \bigl[ D^2(k)\bigr] Q_2
+ \bigl[ D^3(k)\bigr] Q_3 \biggr\}, \\ \label{Prop}
S_\psi(k) = \biggl\{
&\;& \bigl[ S^1(k)\bigr] Q_1
+ \bigl[ S^3(k)\bigr] Q_2
+ \bigl[ S^3(k)\bigr] Q_3 \biggr\}, \nonumber\\
S_\chi(k) = \biggl\{
&\;& \bigl[ S^1(k)\bigr] Q_1
+ \bigl[ S^2(k)\bigr] Q_2
+ \bigl[ S^3(k)\bigr] Q_3 \biggr\}. \nonumber
\end{eqnarray}
Here,
\begin{eqnarray}
&\;& \Delta^1_{\mu\nu}(k) = {\epsilon_{\mu\nu\rho} k^\rho \over \kappa k^2} + {\xi k_\mu k_\nu\over k^4}, \nonumber\\
&\;& \Delta^2_{\mu\nu}(k) = {M_W
(\delta_{\mu\nu} - k_\mu k_\nu/k^2) + \epsilon_{\mu\nu\rho} k^\rho
\over \kappa(k^2 + M_{W}^2)}
+ {\xi k_\mu k_\nu\over k^2(k^2 + {1\over 2} \xi |\vp|^2) }, \nonumber\\
&\;& \Delta^3_{\mu\nu}(k) = {M_Z
(\delta_{\mu\nu} - k_\mu k_\nu/k^2) + \epsilon_{\mu\nu\rho} k^\rho
\over \kappa(k^2 + M_{Z}^2)} + {\xi k_\mu k_\nu\over k^2\bigl[k^2 + {(n-1)\over n} \xi |\vp|^2\bigr] }; \nonumber\\
&\;& D^1(k) =  {1\over (k^2 + {1\over 2} \xi |\vp|^2)}, \nonumber\\
&\;& D^2(k) =  {1 \over (k^2 + M_Z^2)}, \\
&\;& D^3(k) =  {1 \over \bigl[k^2 + {(n-1) \over n} \xi |\vp|^2 \bigr]}; \nonumber\\
&\;& S^1(k) =  {1\over (i\gamma\cdot k - M_W)}, \nonumber\\
&\;& S^2(k) =  {1\over (i\gamma\cdot k + M_Z)}, \nonumber\\
&\;& S^3(k) =  {1\over (i\gamma\cdot k - M_Z)}, \nonumber
\end{eqnarray}
and $M= \kappa g^2, M_{W} = |\vp|^2/(2\kappa),$ and $M_{Z} =
(n-1)|\vp|^2/(n\kappa).$

To determine the renormalization of the Chern-Simons coefficient,
it is sufficient to calculate the parity odd part of the vacuum
polarization. The three relevant diagrams  are shown in Fig. 1:
one with a gluon loop, one with a gluon-Higgs loop and one with a
fermion loop \cite{PV}. After some algebra, we see that the vacuum
polarization can be decomposed into three parts:
\begin{eqnarray}
\bigl[\Pi^{mn}_{\mu\nu}(p) \bigr]_{odd} =
\epsilon_{\mu\nu\rho}p_\rho \biggl\{ \Pi_1(p^2) \delta_{mn}
&\;& + \Pi_2(p^2)\bigl[(\hvp^\dagger T^mT^n\hvp)+(\hvp^\dagger T^nT^m\hvp) \bigr]
\nonumber\\
&\;& \qquad \qquad
+ \Pi_3(p^2) (\hvp^\dagger T^m \hvp) (\hvp^\dagger T^n \hvp) \biggr\}.
\label{PI}
\end{eqnarray}
Since the two would be Chern-Simons terms only contribute to
$\Pi_2(0)$ and $\Pi_3(0)$, we only need to calculate $\Pi_1(0)$ to
find the correction to the Chern-Simons coefficient. In the Landau
gauge,
\begin{equation}
\Pi_1(p) = \Pi_B(p) + 2\Pi_F(p),
\end{equation}
with
\begin{eqnarray}
\Pi_B(p) =
&\;& \int {d^3k\over (2\pi)^3}
\biggl\{ {\bigl[k^2 p^2 -(k\cdot p)^2 \bigr]
\over p^2 (k^2+M_{W}^2) \bigl[(k-p)^2+M_{W}^2\bigr] }\biggr\} \biggl\{
{-M_{W}\over 2(k-p)^2} + {-M_{W}\over 2(k)^2} \biggr\} \nonumber\\
&\;& + \int {d^3k\over (2\pi)^3}
\biggl\{ {- M_W(k\cdot p)\over p^2(k^2+M_{W}^2)(k-p)^2}
+ {M_{W} \bigl[-k^2p^2 - (k\cdot p)^2 + 2k^2(k\cdot p) \bigr]
\over p^2 k^2(k^2+M_{W}^2)(k-p)^2} \biggr\}; \\
\Pi_F(p) =
&\;& \int {d^3k\over (2\pi)^3}
\biggl\{ {M_W \over (k^2 + M_{W}^2)((k-p)^2+M_{W}^2)} \biggr\}.\nonumber
\end{eqnarray}
In the zero momentum limit,
\begin{eqnarray}
&\;& \Pi_B(0) = {-\kappa\over 4\pi|\kappa|}, \\
&\;& \Pi_F(0) = {\kappa\over 8\pi|\kappa|}.  \nonumber
\end{eqnarray}
By throwing away $\phi_1$ and $\chi$, we can also obtain the
correction for $N=2$ supersymmetric Chern-Simons Higgs theories.
In sum, the corrections are
\begin{eqnarray}
&\;& \Delta\kappa_{\rm N=3} = 0, \nonumber\\
&\;& \Delta\kappa_{\rm N=2} = {-\kappa\over 8\pi|\kappa|}. \\
\end{eqnarray}
Both the results are identical to those in the symmetric phase.
Therefore, the degeneracy between the symmetric and asymmetric
vacua is preserved as we have expected for supersymmetric
theories.  This is confirmed by calculating the effective
potential of $\phi_2$.

The situation is quite different, if we introduce the Yang-Mills term as a ultraviolet regulator. From the result in Ref. \cite{Nahill}, we have
\begin{equation}
\Pi_B(p) = {(n-1)\over 2}\Pi^{Ia}(p) + {1\over 2}\Pi^{Ib}(p),
\label{YM}
\end{equation}
in the Landau gauge.  Here,
\begin{eqnarray}
\Pi^{Ia}(p) =
&\;& \int {d^3k\over (2\pi)^3}
\biggl\{ {M\bigl[k^2 p^2 -(k\cdot p)^2 \bigr]
\bigl[ 4M^2 + 10k^2-10k\cdot p + 8p^2 \bigr] \over
p^2 k^2(k^2+M^2)(k-p)^2\bigl[ (k-p)^2+M^2 \bigr]}\biggr\} \nonumber\\
+ &\;& \int {d^3k\over (2\pi)^3}
\biggl\{ {M\bigl[ -2k^2 p^2 -2(k\cdot p)^2 + 4p^2(k\cdot p) \bigr] \over
p^2 k^2(k^2+M^2)(k-p)^2} \biggr\}, \nonumber\\
\Pi^{Ib}(p) =
&\;& \int {d^3k\over (2\pi)^3}
\biggl\{ {M\bigl[k^2 p^2 -(k\cdot p)^2 \bigr] \over p^2
(k^2+M_{W^+}^2)(k^2+M_{W^-}^2)
\bigl[(k-p)^2+M_{W^+}^2\bigr] \bigl[(k-p)^2+M_{W^-}^2\bigr] }\biggr\}\nonumber\\
&\;& \qquad\quad \times \biggl\{ 6M^2 + {(k^2 + M_{W^+} M_{W^-})
\bigl[- M^2 + 8k^2 - 4k\cdot p + 4p^2\bigr]\over k^2} \nonumber\\
&\;& \qquad\qquad + {\bigl[(k-p)^2 + M_{W^+} M_{W^-}\bigr]
\bigl[-M^2 + 8k^2 - 12k\cdot p + 8p^2 \bigr] \over (k-p)^2 } \label{PiI}\\
&\;& \qquad\qquad + {(k^2 + M_{W^+} M_{W^-})\bigl[(k-p)^2 + M_{W^+} M_{W^-} \bigr]
\bigl[-6k^2 + 6k\cdot p -4p^2 \bigr] \over k^2(k-p)^2 } \biggr\} \nonumber\\
+ &\;& \int {d^3k\over (2\pi)^3}
\biggl\{ {-2M(k\cdot p)\bigl[M_{W^+} M_{W^-} + 2k^2 - 2p^2 \bigr] \over
p^2 (k^2 + M_{W^+}^2)(k^2+M_{W^-}^2)(k-p)^2} \nonumber\\
&\;& \qquad\qquad + {M(k^2 + M_{W^+} M_{W^-})
\bigl[-2k^2p^2 - 2(k\cdot p)^2 + 4k^2(k\cdot p) \bigr]
\over p^2 k^2(k^2 + M_{W^+}^2)(k^2+M_{W^-}^2)(k-p)^2} \biggr\}.  \nonumber
\end{eqnarray}
They come from the unbroken and broken sectors, respectively.  In the zero momentum limit, we see
\begin{eqnarray}
&\;& \Pi^{Ia}(0) = {\kappa\over 2\pi|\kappa|}, \nonumber\\
&\;& \Pi^{Ib}(0) = 0.  
\end{eqnarray}
It is obvious that taking the limit that $g\to \infty$ does not change the above result.

For $n\ge 3,$ there is remaining gauge symmetry and
\begin{eqnarray}
&\;& \Pi_B(0) = {(n-1)\kappa\over 4\pi|\kappa|}.
\end{eqnarray}
Consequently,
\begin{eqnarray}
&\;& \Delta\kappa_{\rm N=3} = {n\kappa\over 4\pi|\kappa|}, \nonumber\\
&\;& \Delta\kappa_{\rm N=2} = {(n-1/2)\kappa\over 4\pi|\kappa|}.
\end{eqnarray}
Again, both the above results are identical to those in the symmetric phase. In the $SU(2)$ case the gauge symmetry is completely broken, and there is no such a thing as unbroken part in the Higgs phase.  As a result, the first terms in Eq.~(\ref{YM}) should not have been there.  Since $\Pi^{Ib}(0) = 0$ and thus the bosonic part is vanishing, the quantum correction to the Chern-Simons coefficient comes from the fermionic part only and is
\begin{eqnarray}
&\;& \Delta\kappa_{\rm N=3} = {\kappa\over 4\pi|\kappa|}, \nonumber\\
&\;& \Delta\kappa_{\rm N=2} = {\kappa\over 8\pi|\kappa|}.
\end{eqnarray}
Both results are different from those in the symmetric phase. This
indicates that the supersymmetry is broken when the gauge group is
fundamental $SU(2)$. Since the Yang-Mills term itself does not
respect supersymmetry, it is hardly surprising.  The confusing
part is why this happens only for the $SU(2)$ case.  One possible
way to clarify the above confusion is to do a derivative expansion
type of calculation as in Ref.\cite{CHTh}.

The results that the quantum correction to the Chern-Simons
coefficient in supersymmetric Chern-Simons Higgs theories are
identical in the symmetric and Higgs phases is interesting and
have important implication.  It is well-known that non-abelian
self-dual Chern-Simons Higgs theories with the Higgs in the
adjoint representation have rich vacuum structure.  It has been
quite a challenge to verify that the quantum correction to the
Chern-Simons coefficient is quantized in these systems. If the
results obtained above can be generalized to the adjoint
representation, we can calculate the quantum correction in
self-dual Chern-Simons Higgs theories by calculating the fermionic
part in the corresponding supersymmetric Chern-Simons Higgs
theories.  Finally, although the calculation done here is only for supersymmtric Chern-Simons Higgs theories, we believe the results also apply to supersymmetric Yang-Mills Chern-Simons Higgs theories based on our experience from Ref.\cite{SUYMCS}.
\vskip1cm
\noindent{\bf Acknowledgments}

This work is supported in part by the National Science Council of
R.O.C. under grant No. NSC88-2112-M-032-003.

\vfill\eject

\begin{figure}[h]
\includegraphics{sdcscffig.ps}
\vskip 6.5cm
\end{figure}


\end{document}